\documentclass[showpacs,twocolumn,prl]{revtex4}

\usepackage{graphicx}
\usepackage{dcolumn}
\usepackage{bm}
\usepackage{amsmath}

\begin{document}

\title{Single-parameter non-adiabatic quantized charge pumping}

\author{B. Kaestner}
\email[Electronic address: ]{Bernd.Kaestner@ptb.de}
\affiliation{
		Physikalisch-Technische Bundesanstalt, Bundesallee 100, 38116 Braunschweig, Germany
}

\author{V. Kashcheyevs}
\affiliation{
		Department of Physics, Ben Gurion University of the Negev, Beer Sheva 84105, Israel
}
\affiliation{
		Institute for Solid State Physics, University of Latvia, Riga LV-1063, Latvia
}

\author{S. Amakawa}
\affiliation{
		Integrated Research Institute, Tokyo Institute of Technology, 4259-R2-17 Nagatsuta, 
		Midori-ku, Yokohama 226-8503, Japan
}

\author{L. Li}
\affiliation{
 		Sichuan Normal University, No. 5, Jing'an Road, Chengdu 610066, Sichuan, China
}

\author{M. D. Blumenthal, T. J. B. M. Janssen}
\affiliation{
		Cavendish Laboratory, University of Cambridge, Cambridge CB3 0HE, UK
}
\affiliation{
		National Physical Laboratory, Hampton Road, Teddington TW11 0LW, UK
}

\author{G. Hein, K. Pierz, T. Weimann, U. Siegner, and H. W. Schumacher}
\affiliation{
		Physikalisch-Technische Bundesanstalt, Bundesallee 100, 38116 Braunschweig, Germany.
}

\date{\today}

\begin{abstract}

Controlled charge pumping in an AlGaAs/GaAs gated nanowire by single-parameter modulation is studied experimentally and theoretically. Transfer of integral multiples of the elementary charge per modulation cycle is clearly demonstrated. A simple theoretical model shows that such a quantized current can be generated via loading and unloading of a dynamic quasi-bound state. It demonstrates that non-adiabatic blockade of unwanted tunnel events can obliterate the requirement of having at least two phase-shifted periodic signals to realize quantized pumping. The simple configuration without multiple pumping signals might find wide application in metrological experiments and quantum electronics.

\end{abstract}


\maketitle

An important milestone in the study of single electron transport is the closure of the quantum metrological triangle for frequency, dc current, and dc voltage \cite{likharev1985}. Dc voltage is currently realized from the frequency standard through the Josephson effect.  Dc current can then be derived using the quantum Hall effect.  Direct realization of dc current from frequency is the currently missing side of the triangle.  The closure of the quantum metrological triangle provides a test whether the fundamental constants really appear the same in these different systems \cite{NIU1990}. The results of this kind of experiment will also impact on a future system of units which might be based on fundamental constants \cite{mills2006}.

A current source relevant for the above experiments must produce at least nanoampere currents to be measurable with sufficient accuracy.  The electron pump based on arrays of Coulomb blockaded quantum dots (see \cite{esteve1992} for a review) or quantum interference \cite{thouless1, NIU1990, Levinson2001} is one class of devices being investigated with respect to metrological relevance \cite{kautz1999, lotkhov2001, blumenthal2007a}. Electron pumps are typically driven by multiple radio frequency (rf) signals with a well maintained phase relationship, producing a {\em quantized} current, i.e.\ limited to certain values according to $I = - n e f$ (with $n = 1,2,3 \dots$, $e$ the negative elementary charge and $f$ the driving frequency). Usually, the accuracy in $I$ degrades with increasing $f$, which has so far prevented the generation of sufficiently accurate nanoampere currents. An alternative, but challenging task would be the parallelization of pumps driven at intermediate frequencies.
Here, pumps requiring only a {\em single} rf signal would fundamentally reduce the complexity in the parallelization of such devices.
However, electron pumps driven by only one gate \cite{Jalil1998, KouwenhovenB91, Tsukagoshi3PBI, Altebaeumer2PBI} have so far not experimentally demonstrated the generation of quantized current. In addition, most models of quantized pumping \cite{thouless1, Levinson2001,VKAAOE03res,wohlman2002,maksym1,pothier1PBI} have assumed at least two parameters modulated out phase, which may be motivated by the fact that in the adiabatic limit a single periodic perturbation cannot determine the direction of the current \cite{moskalets2002B}.

In this paper we address this issue and report on the first experimental realization of quantized charge pumping in which only one gate is modulated. 
We demonstrate on a transparent quantum model that the observed current quantization can be explained via a simple loading and unloading mechanism of a dynamic quasi-bound state. Previously, such a mechanism was suggested \cite{VKAAOE03res} to explain the theoretically predicted quantization in two-parameter adiabatic quantum pumps \cite{Levinson2001,wohlman2002}, where the necessary switch in spatial asymmetry between loading and unloading is ensured by a phase shift between two periodic signals. The novelty of our scenario is that an effective phase-shift is intrinsically generated by non-adiabatic blockade of tunneling, for which the modulation of a single parameter is sufficient. Thus non-adiabaticity in our device not only plays a constructive role~\cite{maksym1} but is essential for the quantized pump to work.

\begin{figure}
\includegraphics{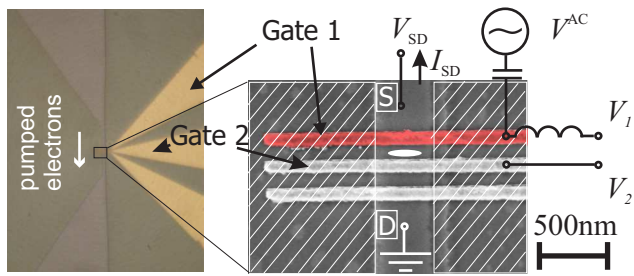}
\caption{\label{fig:device} (Color online) Picture of the device (left) with TiAu finger gates over the etched channel. In the SEM picture (right) bias and gate voltages are indicated, showing the gate colored in red as being modulated. The source (S) and drain (D) reservoirs are indicated.
The hatched regions are depleted of the 2D electron gas, defining a wire of 500$\,$nm in width. A quasi-bound state is formed between Gates 1 and 2, as indicated by the white ellipse. The direction of the pumped electrons is indicated by the white arrow on the left. The lowest gate is not in use in this experiment.}
\end{figure}

\begin{figure}
\includegraphics{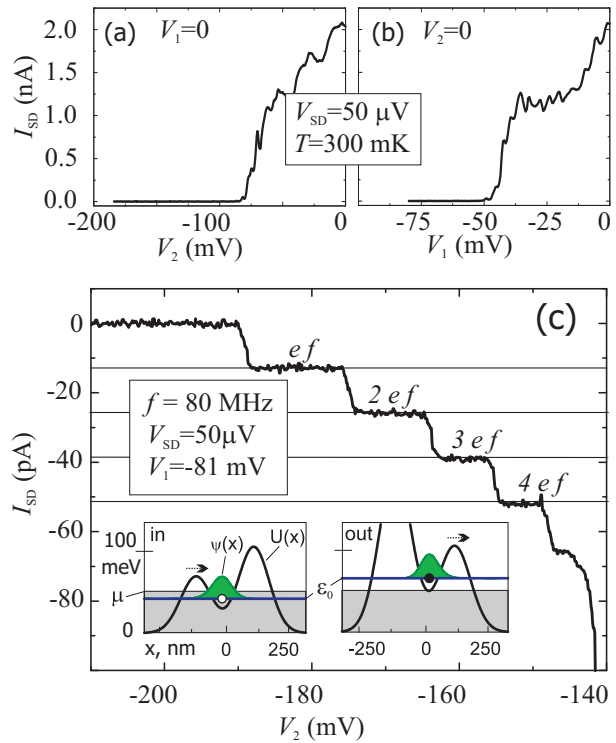}
\caption{\label{fig:Plateaus} Gate characteristics shown in (a) and (b). In (c) $I_\mathrm{SD}$ is plotted when rf modulation is applied to Gate 1. 
Snapshots of the time dependent potential $U$ during loading and unloading of a single electron, are shown in the insets. Calculated $U$ and the wavefunction of the relevant transport state $\psi$ correspond to the calculation presented in Fig.~\ref{fig:timedep}.}
\end{figure}

The device is realized by two 100$\,$nm-wide metallic finger gates crossing a wire etched in an $n$-type AlGaAs heterostructure, as shown in Fig.~\ref{fig:device}. The lithographic width of the wire is 500$\,$nm and the distance between the gate centers is 250$\,$nm. All measurements were performed in a $^3$He cryostat with a base temperature of 300$\,$mK. Figs.~\ref{fig:Plateaus}a and b show the characteristics of the two metallic finger gates. The source-drain current $I_\mathrm{SD}$ is plotted as a function of gate voltages $V_1$ and $V_2$ applied to Gate 1 and 2, respectively. The wire is biased with $V_\mathrm{SD} = +50\,\mu$V. 
The pinch-off voltages of Gates 1 and 2 are $-50\,$mV and $-80\,$mV, respectively, while the other gate is grounded.

For quantized charge pumping the device is operated in the following way: Both gates are driven beyond pinch-off by static voltages of $V_1 = - 81\,$mV and $V_2 = -140\,$mV. 
A sine wave modulation with frequency $f = 80\,$MHz is superimposed onto $V_1$. 
The applied rf power of $13.5\,$dBm corresponds to an amplitude at Gate 1 of approximately $\pm 75\,$mV, as estimated by conductance measurements. 
For values of $V_2 < -140\,$mV electrons are pumped from source to drain, i.e. $I_\mathrm{SD} < 0$, as shown in Fig.~\ref{fig:Plateaus}c. 
Four clear plateaus are observed as $V_2$ is varied from $-145 \dots -190\,$mV. The plateaus are separated by $\Delta I = 12.8\,$pA as expected when the number of transfered electrons per cycle changes by one.
Therefore we conclude that in this configuration up to 4 electrons can be transfered in one cycle, depending on the value of $V_2$.
The current can be generated in the unbiased device, but here a bias voltage, $V_\mathrm{SD} = +50\,\mu$V, was applied opposite to the pumping direction to demonstrate the robustness of the quantized pumping effect.
While the length of the plateaus decreases with increasing $n$, no degrading of the plateau flatness is observed. Thus, the number of pumped electrons is stable over a controllable parameter range. The current value for $n = 4$ was determined around the center of the plateau to be $I_\mathrm{SD} = (- 52 \pm 1)\,$pA, where the accuracy was limited by the measurement setup, not by the device. The theoretical value of $4 e f = - 51.3\,$pA lies well within the measurement accuracy. Quantized current has been measured for $V_1 = -40$ to $-95\,$mV. Similar results were obtained for a range of modulation frequencies from 75 to 85$\,$MHz, with the corresponding plateau values.

For a quantitative theoretical analysis of the quantization mechanism we consider a simple quantum model of
non-interacting electrons confined in a one-dimensional wire and subjected to a time-dependent double-barrier potential plotted as inset in Fig.~\ref{fig:Plateaus}c
\begin{align} \label{eq:Potential}
  U(x,t) & = U_1(t) \, e^{-(x+x_0)^2/w^2} + U_2 \, e^{-(x-x_0)^2/w^2}
\end{align}
with a harmonically oscillating left barrier, $U_1(t) = U_1^{\text{dc}}- U_1^{\text{ac}} \cos (2 \pi f t)$. The
boundary conditions are defined by a Fermi distribution of electrons coming from the left, $f_\mathrm{F}(\mu+e V_\mathrm{SD})$, and from the right, $f_\mathrm{F}(\mu)$, where $\mu$ is the electrochemical potential of the drain. Standard parabolic dispersion is taken for the wire assuming bulk GaAs effective electron mass $m^{\ast} =0.067 m_e$.

Full statistics of the stationary state in this model, including the dc current and the Fano factor, can in principle be obtained by solving the corresponding Floquet scattering problem \cite{moskalets2002B,kohler2004}. However, the high number of excited side-bands in the vicinity of the adiabatic limit renders such a calculation impractical. In order to proceed with the calculation, we restrict the parameters such that at all times there exists at least one quasi-bound state in the potential well formed between the barriers. 

The instantaneous energy level $\epsilon_0(t)$ and its broadenings due to tunneling coupling to the left, $\Gamma_\mathrm{L}(t)$, and to the right, $\Gamma_\mathrm{R}(t)$
for the lowest of these states are obtained numerically by solving the frozen-time scattering problem and
approximating the corresponding resonance with a Breit-Wigner formula.

The other quasi-bound states can be ignored if the gap from the lowest state,  $\Delta \epsilon \equiv \epsilon_1 - \epsilon_0$, is sufficiently large, $\Delta \epsilon> \mu-\epsilon_0, h f, k_B T$. It has been shown in Ref.~\onlinecite{VKAAOE03res} (preprint version) that exact results for \emph{adiabatic} ($f \to 0$) pumping via a single resonance can be accurately approximated for $\Gamma \ll k_B T$ by solving a simple rate equation for the level occupation probability $P(t)$:
\begin{align} \label{eq:rateeq}
  \hbar \dot{P} = (\Gamma_\mathrm{L}+\Gamma_\mathrm{R}) \left [ f_\mathrm{F}(\epsilon_0) - P \right ]
\end{align}
(henceforth we consider pure pumping only, $V_\mathrm{SD}=0$).

We shall assume Eq.~\eqref{eq:rateeq} to hold also in the non-adiabatic regime as long as the only characteristic energy scales allowed to be less than $h f$ are the tunneling widths $\Gamma_L$ and $\Gamma_R$. The average dc current, $I = - e N f$ (where $N$ is a real number), is then easily calculated by separating the left and right contributions to the full tunneling current in the r.h.s.~of Eq.~\eqref{eq:rateeq}. Physically this assumption corresponds to neglecting the dynamics of hot electrons and holes outside the double-barrier structure.

\begin{figure}
  \includegraphics[width=8cm]{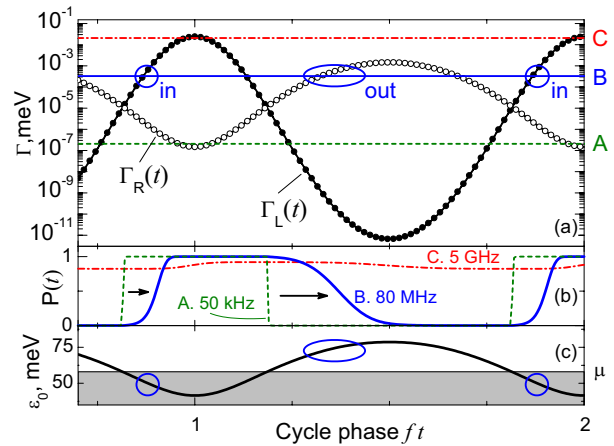}
  \caption{(Color online)(a) Instantaneous values of the tunneling broadenings $\Gamma_{L}(t)$ and $\Gamma_{R}(t)$ for the lowest quasi-localized energy level $\epsilon_0(t)$. The left barrier oscillates harmonically with an amplitude $U_1^{\text{ac}}=100 \, \text{meV}$ around the value $U_1^{\text{dc}} = 160 \,  \text{meV}$, the right barrier is fixed at $U_2=120 \, \text{meV}$. Horizontal lines $A$, $B$ and $C$ mark the energy quanta $h f$ for three representative frequencies. (b) Corresponding level occupation probability $P(t)$ at the selected frequencies. (c) Time evolution of the lowest quasi-localized energy $\epsilon_0(t)$. The blue ellipses indicate regions of charge exchange in regime B.
Model parameters are $2 \, x_0=250 \, \text{nm}$, $w=95 \, \text{nm}$, $\mu=58 \, \text{meV}$, $T=300  \, \text{mK}$.
  \label{fig:timedep}}
\end{figure}

\begin{figure}
  \includegraphics[width=8cm]{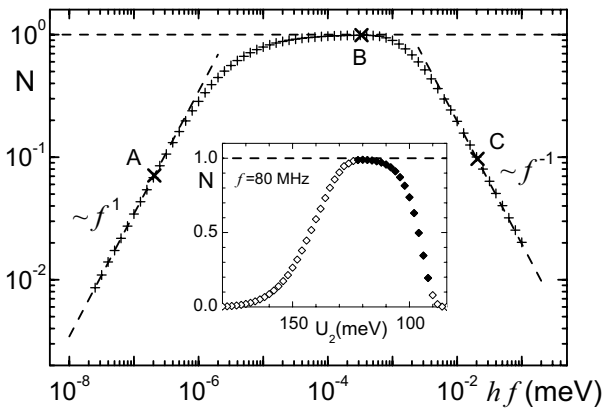}
  \caption{Calculated average number of pumped electrons per cycle
  as a function of the driving frequency. Model parameters are
  the same as in Fig.~\ref{fig:timedep}. Inset:
  number of pumped electrons at the experimental 
  frequency $f=80$ MHz (point B of the main plot) as a 
  function of the fixed barrier height $U_2$. Open symbols indicate 
  points in the parameter space at which the total width 
  $\Gamma_L+\Gamma_R$ of the quasi-bound state briefly
  exceeds $k_B T$ at the extreme values of $U_1(t)$.
  \label{fig:freqdep}}
\end{figure}

Results of our calculations, summarized in Figs.~\ref{fig:timedep} and \ref{fig:freqdep} and discussed below, reveal three physically different transport regimes: (A) small next-order non-adiabatic corrections \cite{wohlman2002}, $N \propto f$, to the symmetry-dictated \cite{moskalets2002B} adiabatic limit $N(f\to0)=0$; (B)  current quantization, $N \approx 1$, as achieved in the present experiment; and (C) crossover to approximately frequency-independent current, $N \propto f^{-1}$, at high frequencies. Typical evolution of 
the tunneling couplings during one cycle is shown in Fig.~\ref{fig:timedep}a. Coupling to the left $\Gamma_L$ changes exponentially because of left barrier height $U_1$ modulation, while $\Gamma_R$ changes mainly due to oscillations in $\epsilon_0$.

The qualitative behavior of our pumping model is determined by the competition between tunneling and non-adiabaticity. At finite frequency $f$, the transport becomes blocked as soon as the corresponding tunneling coupling $\Gamma$ is less than a characteristic scale of order $h f$ \cite{Flensberg1999,moskalets2002B}.
Three representative values of $h f$ are indicated in Fig.~\ref{fig:timedep}a corresponding to the scenarios A, B, and C.

In the weakly non-adiabatic regime (A), the tunneling between the local level and at least one of the two reservoirs is quick enough to keep the occupation probability close to instantaneous equilibrium, $P(t) \approx f_F(\epsilon_0(t) )$ (see curve A Fig.~\ref{fig:timedep}b). Transitions of $P(t)$ between $0$ and $1$ (charge loading and unloading of Ref.~\onlinecite{VKAAOE03res}) in this case appear as sharp steps at the time moments when the transport level crosses the Fermi energy, $\epsilon_0(t)=\mu$, which can be obtained from Fig.~\ref{fig:timedep}c. Because of only one pumping parameter, the ratio $\Gamma_L/\Gamma_R$ is the same at these two charge transfer points, and the net pumped charge remains close to zero. 

A qualitative difference from the nearly-adiabatic regime (A) is observed once the frequency is high enough to ensure that in a certain part of the period tunneling to both sides is effectively switched off (regime B). Then the loading (``in'') and unloading (``out'') processes are delayed with respect to crossing of the Fermi level. The reason for the delay is that the tunneling rates, $\Gamma/\hbar$, first have to grow sufficiently large to allow for charge exchange with the leads (see regions marked by ellipses in Fig.~\ref{fig:timedep}a and c, and corresponding potential snapshots in Fig~\ref{fig:Plateaus}). This delay is clearly visible in the
plots of $P(t)$ as marked by arrows in Fig.~\ref{fig:timedep}b. As a result, the barrier asymmetry ratio $\Gamma_L/\Gamma_R$ has time to become $\gg 1$ for loading, and $\ll 1$ for unloading as required for nearly integer charge transfer from left to right in one cycle.

Frequency increase beyond the optimal range of operation would lead to incomplete loading and unloading, and as a result reduce the accuracy. This reduction is similar to the degrading role of non-adiabaticity discussed in Ref.~\cite{Flensberg1999}. Eventually, in the high-frequency regime (C) of our model the tunneling events become rare on the time-scale of a single period, and the occupation probability $P(t)$ approaches a constant time-averaged value $\overline{(\Gamma_{L}+\Gamma_{R}) f_F(\epsilon_0)}/{(\overline{\Gamma_{L}+\Gamma_{R}})}$.

The results of the calculation of the average charge transfer per cycle are shown in Fig.~\ref{fig:freqdep} for a range of frequencies. Since the precise shape and magnitude of the screened gate potential inside the channel are not known, the barrier heights and width in Eq.~\eqref{eq:Potential} have been optimized once for Fig.~\ref{fig:timedep} and Fig.~\ref{fig:freqdep} to give the best-quantized plateau at the experimental frequency of $f=80$ MHz while remaining within the validity range of our rate-equation-based calculation. 
Changing the potential on Gate 2 reveals the current quantization step shown in the inset in Fig.~\ref{fig:freqdep} (compare to experiment in Fig.~\ref{fig:Plateaus}). The current drops as  $U_2$ is decreased because the right barrier becomes leaky and allows loading from the drain. In the experiment, this leakage determines the maximal range over which plateaus can be measured, i.e.\ for voltages $V_2 < -145\,$mV.

In conclusion, we have demonstrated quantized charge pumping through a nanowire using one modulated gate at 80 MHz and one gate with a fixed gate voltage to select the number of charges pumped per cycle. 
The possibility to produce quantized current via single-barrier modulation significantly reduces the complexity in 
operating a large number of {\em independent} devices in parallel with a single rf signal.
The principle of operation is analysed in a simple quantum mechanical model. Besides providing a transparent framework for describing the quantization mechanism it shows the essential role of non-adiabaticity for the pump to work. This understanding together with the technical advantages of single-parameter pumping could open the door for an accurately quantized, large-current source as needed for fundamental experiments in metrology and quantum electronics.

\begin{acknowledgments}
The authors acknowledge fruitful discussions with T. Altebaeumer, S. Kohler, and F. J. Kaiser. V. K. acknowledges financial support from German-Israeli Project Cooperation (DIP), Israel Science Foundation and Latvian Council of Science.
\end{acknowledgments}

\bibliography{literatureJabRef}

\begin{thebibliography}{20}
\expandafter\ifx\csname natexlab\endcsname\relax\def\natexlab#1{#1}\fi
\expandafter\ifx\csname bibnamefont\endcsname\relax
  \def\bibnamefont#1{#1}\fi
\expandafter\ifx\csname bibfnamefont\endcsname\relax
  \def\bibfnamefont#1{#1}\fi
\expandafter\ifx\csname citenamefont\endcsname\relax
  \def\citenamefont#1{#1}\fi
\expandafter\ifx\csname url\endcsname\relax
  \def\url#1{\texttt{#1}}\fi
\expandafter\ifx\csname urlprefix\endcsname\relax\def\urlprefix{URL }\fi
\providecommand{\bibinfo}[2]{#2}
\providecommand{\eprint}[2][]{\url{#2}}

\bibitem[{\citenamefont{Likharev and Zorin}(1985)}]{likharev1985}
\bibinfo{author}{\bibfnamefont{K.~K.} \bibnamefont{Likharev}} \bibnamefont{and}
  \bibinfo{author}{\bibfnamefont{A.~B.} \bibnamefont{Zorin}},
  \bibinfo{journal}{J. Low Temp. Phys.} \textbf{\bibinfo{volume}{59}},
  \bibinfo{pages}{347} (\bibinfo{year}{1985}).

\bibitem[{\citenamefont{Niu}(1990)}]{NIU1990}
\bibinfo{author}{\bibfnamefont{Q.}~\bibnamefont{Niu}}, \bibinfo{journal}{Phys.
  Rev. Lett.} \textbf{\bibinfo{volume}{64}}, \bibinfo{pages}{1812}
  (\bibinfo{year}{1990}).

\bibitem[{\citenamefont{Mills et~al.}(2006)\citenamefont{Mills, Mohr, Quinn,
  Taylor, and Williams}}]{mills2006}
\bibinfo{author}{\bibfnamefont{I.~M.} \bibnamefont{Mills}},
  \bibinfo{author}{\bibfnamefont{P.~J.} \bibnamefont{Mohr}},
  \bibinfo{author}{\bibfnamefont{T.~J.} \bibnamefont{Quinn}},
  \bibinfo{author}{\bibfnamefont{B.~N.} \bibnamefont{Taylor}},
  \bibnamefont{and} \bibinfo{author}{\bibfnamefont{E.~R.}
  \bibnamefont{Williams}}, \bibinfo{journal}{Metrologia}
  \textbf{\bibinfo{volume}{43}}, \bibinfo{pages}{227} (\bibinfo{year}{2006}).

\bibitem[{\citenamefont{Esteve}(1992)}]{esteve1992}
\bibinfo{author}{\bibfnamefont{D.}~\bibnamefont{Esteve}},
  \emph{\bibinfo{title}{Single charge tunneling}} (\bibinfo{publisher}{Plenum
  Press, New York}, \bibinfo{year}{1992}), chap.~\bibinfo{chapter}{4}, pp.
  \bibinfo{pages}{109--137}.

\bibitem[{\citenamefont{Thouless}(1983)}]{thouless1}
\bibinfo{author}{\bibfnamefont{D.~J.} \bibnamefont{Thouless}},
  \bibinfo{journal}{Phys. Rev. B} \textbf{\bibinfo{volume}{27}},
  \bibinfo{pages}{6083} (\bibinfo{year}{1983}).

\bibitem[{\citenamefont{Levinson et~al.}(2001)\citenamefont{Levinson,
  Entin-Wohlman, and W{\"{o}}lfle}}]{Levinson2001}
\bibinfo{author}{\bibfnamefont{Y.}~\bibnamefont{Levinson}},
  \bibinfo{author}{\bibfnamefont{O.}~\bibnamefont{Entin-Wohlman}},
  \bibnamefont{and}
  \bibinfo{author}{\bibfnamefont{P.}~\bibnamefont{W{\"{o}}lfle}},
  \bibinfo{journal}{Physica A} \textbf{\bibinfo{volume}{302}},
  \bibinfo{pages}{335} (\bibinfo{year}{2001}).

\bibitem[{\citenamefont{Kautz et~al.}(1999)\citenamefont{Kautz, Keller, and
  Martinis}}]{kautz1999}
\bibinfo{author}{\bibfnamefont{R.~L.} \bibnamefont{Kautz}},
  \bibinfo{author}{\bibfnamefont{M.~W.} \bibnamefont{Keller}},
  \bibnamefont{and} \bibinfo{author}{\bibfnamefont{J.~M.}
  \bibnamefont{Martinis}}, \bibinfo{journal}{Phys. Rev. B}
  \textbf{\bibinfo{volume}{60}}, \bibinfo{pages}{8199} (\bibinfo{year}{1999}).

\bibitem[{\citenamefont{Lotkhov et~al.}(2001)\citenamefont{Lotkhov,
  Bogoslovsky, Zorin, and Niemeyer}}]{lotkhov2001}
\bibinfo{author}{\bibfnamefont{S.~V.} \bibnamefont{Lotkhov}},
  \bibinfo{author}{\bibfnamefont{S.~A.} \bibnamefont{Bogoslovsky}},
  \bibinfo{author}{\bibfnamefont{A.~B.} \bibnamefont{Zorin}}, \bibnamefont{and}
  \bibinfo{author}{\bibfnamefont{J.}~\bibnamefont{Niemeyer}},
  \bibinfo{journal}{Appl. Phys. Lett.} \textbf{\bibinfo{volume}{78}},
  \bibinfo{pages}{946} (\bibinfo{year}{2001}).

\bibitem[{\citenamefont{Blumenthal et~al.}(2007)\citenamefont{Blumenthal,
  Kaestner, Li, Giblin, Janssen, Pepper, Anderson, Jones, and
  Ritchie}}]{blumenthal2007a}
\bibinfo{author}{\bibfnamefont{M.~D.} \bibnamefont{Blumenthal}},
  \bibinfo{author}{\bibfnamefont{B.}~\bibnamefont{Kaestner}},
  \bibinfo{author}{\bibfnamefont{L.}~\bibnamefont{Li}},
  \bibinfo{author}{\bibfnamefont{S.}~\bibnamefont{Giblin}},
  \bibinfo{author}{\bibfnamefont{T.~J. B.~M.} \bibnamefont{Janssen}},
  \bibinfo{author}{\bibfnamefont{M.}~\bibnamefont{Pepper}},
  \bibinfo{author}{\bibfnamefont{D.}~\bibnamefont{Anderson}},
  \bibinfo{author}{\bibfnamefont{G.}~\bibnamefont{Jones}}, \bibnamefont{and}
  \bibinfo{author}{\bibfnamefont{D.~A.} \bibnamefont{Ritchie}},
  \bibinfo{journal}{Nature Physics} \textbf{\bibinfo{volume}{3}},
  \bibinfo{pages}{343 } (\bibinfo{year}{2007}).

\bibitem[{\citenamefont{Jalil et~al.}(1998)\citenamefont{Jalil, Ahmed, and
  Wagner}}]{Jalil1998}
\bibinfo{author}{\bibfnamefont{M.~B.~A.} \bibnamefont{Jalil}},
  \bibinfo{author}{\bibfnamefont{H.}~\bibnamefont{Ahmed}}, \bibnamefont{and}
  \bibinfo{author}{\bibfnamefont{M.}~\bibnamefont{Wagner}},
  \bibinfo{journal}{J. Appl. Phys.} \textbf{\bibinfo{volume}{84}},
  \bibinfo{pages}{4617} (\bibinfo{year}{1998}).

\bibitem[{\citenamefont{Kouwenhoven et~al.}(1991)\citenamefont{Kouwenhoven,
  Johnson, {van der Vaart}, {van der Enden}, Harmans, and
  Foxon}}]{KouwenhovenB91}
\bibinfo{author}{\bibfnamefont{L.~P.} \bibnamefont{Kouwenhoven}},
  \bibinfo{author}{\bibfnamefont{A.~T.} \bibnamefont{Johnson}},
  \bibinfo{author}{\bibfnamefont{N.~C.} \bibnamefont{{van der Vaart}}},
  \bibinfo{author}{\bibfnamefont{A.}~\bibnamefont{{van der Enden}}},
  \bibinfo{author}{\bibfnamefont{C.~J. P.~M.} \bibnamefont{Harmans}},
  \bibnamefont{and} \bibinfo{author}{\bibfnamefont{C.~T.} \bibnamefont{Foxon}},
  \bibinfo{journal}{Z. Phys. B} \textbf{\bibinfo{volume}{85}},
  \bibinfo{pages}{381} (\bibinfo{year}{1991}).

\bibitem[{\citenamefont{Tsukagoshi et~al.}(1997)\citenamefont{Tsukagoshi,
  Nakazato, Ahmed, and Gamo}}]{Tsukagoshi3PBI}
\bibinfo{author}{\bibfnamefont{K.}~\bibnamefont{Tsukagoshi}},
  \bibinfo{author}{\bibfnamefont{K.}~\bibnamefont{Nakazato}},
  \bibinfo{author}{\bibfnamefont{H.}~\bibnamefont{Ahmed}}, \bibnamefont{and}
  \bibinfo{author}{\bibfnamefont{K.}~\bibnamefont{Gamo}},
  \bibinfo{journal}{Phys. Rev. B} \textbf{\bibinfo{volume}{56}},
  \bibinfo{pages}{3972} (\bibinfo{year}{1997}).

\bibitem[{\citenamefont{Altebaeumer et~al.}(2001)\citenamefont{Altebaeumer,
  Amakawa, and Ahmed}}]{Altebaeumer2PBI}
\bibinfo{author}{\bibfnamefont{T.}~\bibnamefont{Altebaeumer}},
  \bibinfo{author}{\bibfnamefont{S.}~\bibnamefont{Amakawa}}, \bibnamefont{and}
  \bibinfo{author}{\bibfnamefont{H.}~\bibnamefont{Ahmed}},
  \bibinfo{journal}{Appl. Phys. Lett.} \textbf{\bibinfo{volume}{79}},
  \bibinfo{pages}{533} (\bibinfo{year}{2001}).

\bibitem[{\citenamefont{Kashcheyevs et~al.}(2004)\citenamefont{Kashcheyevs,
  Aharony, and Entin-Wohlman}}]{VKAAOE03res}
\bibinfo{author}{\bibfnamefont{V.}~\bibnamefont{Kashcheyevs}},
  \bibinfo{author}{\bibfnamefont{A.}~\bibnamefont{Aharony}}, \bibnamefont{and}
  \bibinfo{author}{\bibfnamefont{O.}~\bibnamefont{Entin-Wohlman}},
  \bibinfo{journal}{Phys. Rev. B} \textbf{\bibinfo{volume}{69}},
  \bibinfo{pages}{195301} (\bibinfo{year}{2004}), \eprint{cond-mat/0308382v1}.

\bibitem[{\citenamefont{Entin-Wohlman et~al.}(2002)\citenamefont{Entin-Wohlman,
  Aharony, and Levinson}}]{wohlman2002}
\bibinfo{author}{\bibfnamefont{O.}~\bibnamefont{Entin-Wohlman}},
  \bibinfo{author}{\bibfnamefont{A.}~\bibnamefont{Aharony}}, \bibnamefont{and}
  \bibinfo{author}{\bibfnamefont{Y.}~\bibnamefont{Levinson}},
  \bibinfo{journal}{Phys. Rev. B} \textbf{\bibinfo{volume}{65}},
  \bibinfo{pages}{195411} (\bibinfo{year}{2002}).

\bibitem[{\citenamefont{Maksym}(2000)}]{maksym1}
\bibinfo{author}{\bibfnamefont{P.}~\bibnamefont{Maksym}},
  \bibinfo{journal}{Phys. Rev. B} \textbf{\bibinfo{volume}{61}},
  \bibinfo{pages}{4727} (\bibinfo{year}{2000}).

\bibitem[{\citenamefont{Pothier et~al.}(1992)\citenamefont{Pothier, Lafarge,
  Urbina, Esteve, and Devoret}}]{pothier1PBI}
\bibinfo{author}{\bibfnamefont{H.}~\bibnamefont{Pothier}},
  \bibinfo{author}{\bibfnamefont{P.}~\bibnamefont{Lafarge}},
  \bibinfo{author}{\bibfnamefont{C.}~\bibnamefont{Urbina}},
  \bibinfo{author}{\bibfnamefont{D.}~\bibnamefont{Esteve}}, \bibnamefont{and}
  \bibinfo{author}{\bibfnamefont{M.~H.} \bibnamefont{Devoret}},
  \bibinfo{journal}{Europhys. Lett.} \textbf{\bibinfo{volume}{17}},
  \bibinfo{pages}{249} (\bibinfo{year}{1992}).

\bibitem[{\citenamefont{Moskalets and B\"uttiker}(2002)}]{moskalets2002B}
\bibinfo{author}{\bibfnamefont{M.}~\bibnamefont{Moskalets}} \bibnamefont{and}
  \bibinfo{author}{\bibfnamefont{M.}~\bibnamefont{B\"uttiker}},
  \bibinfo{journal}{Phys. Rev. B} \textbf{\bibinfo{volume}{66}},
  \bibinfo{pages}{205320} (\bibinfo{year}{2002}).

\bibitem[{\citenamefont{Kohler et~al.}(2005)\citenamefont{Kohler, Lehmann, and
  H\"anggi}}]{kohler2004}
\bibinfo{author}{\bibfnamefont{S.}~\bibnamefont{Kohler}},
  \bibinfo{author}{\bibfnamefont{J.}~\bibnamefont{Lehmann}}, \bibnamefont{and}
  \bibinfo{author}{\bibfnamefont{P.}~\bibnamefont{H\"anggi}},
  \bibinfo{journal}{Phys. Rep.} \textbf{\bibinfo{volume}{406}},
  \bibinfo{pages}{379} (\bibinfo{year}{2005}).

\bibitem[{\citenamefont{Flensberg et~al.}(1999)\citenamefont{Flensberg, Niu,
  and Pustilnik}}]{Flensberg1999}
\bibinfo{author}{\bibfnamefont{K.}~\bibnamefont{Flensberg}},
  \bibinfo{author}{\bibfnamefont{Q.}~\bibnamefont{Niu}}, \bibnamefont{and}
  \bibinfo{author}{\bibfnamefont{M.}~\bibnamefont{Pustilnik}},
  \bibinfo{journal}{Phys. Rev. B} \textbf{\bibinfo{volume}{60}},
  \bibinfo{pages}{R16291} (\bibinfo{year}{1999}).

\end{thebibliography}

\end{document}